\documentstyle[twoside,epsfig,fleqn,espcrc2]{article}

\newcommand{\AmS}{{\protect\the\textfont2
  A\kern-.1667em\lower.5ex\hbox{M}\kern-.125emS}}

\title{Solar panels as air Cherenkov detectors for \\
 extremely high energy cosmic rays}

\author{S.~Cecchini $^{a,d}$, I.~D'Antone $^a$, L.~Degli Esposti $^a$,
 G.~Giacomelli $^a$, M.~Guerra $^b$, I.~Lax $^a$, \\
 G.~Mandrioli $^a$, A.~Parretta $^c$, A.~Sarno $^c$, R.~Schioppo $^b$,
 M.~Sorel $^a$\thanks{
Corresponding author: SOREL@BO.INFN.IT}, M.~Spurio $^a$ \bigskip \\
\noindent $^a$ Dipartimento di Fisica dell'Universit\`{a} di Bologna and
 INFN Sezione di Bologna, 40126 Bologna, Italy \\
\noindent $^b$ ENEA, Area Sperimentale di Monte Aquilone, 71043 Manfredonia
 (FG), Italy \\
\noindent $^c$ ENEA, Centro Ricerche, 80055 Portici
 (NA), Italy \\     
\noindent $^d$ Istituto TESRE/CNR, 40129 Bologna, Italy}

\begin{document}

\begin{abstract}
\indent Increasing interest towards the observation of the highest
 energy cosmic rays has motivated the development
 of new detection techniques. The properties of the Cherenkov
 photon pulse emitted in the atmosphere by these very rare
 particles indicate low-cost semiconductor detectors as good
 candidates for their optical read-out. \\
\indent The aim of this paper is to evaluate the viability
 of solar panels for this purpose. The experimental
 framework resulting from measurements performed with suitably-designed
 solar cells
 and large conventional photovoltaic areas is presented. A discussion
 on the obtained and achievable sensitivities follows. 
\end{abstract}

\maketitle
\section{INTRODUCTION}
\begin{figure}[htb]
\centerline{
\epsfig{file=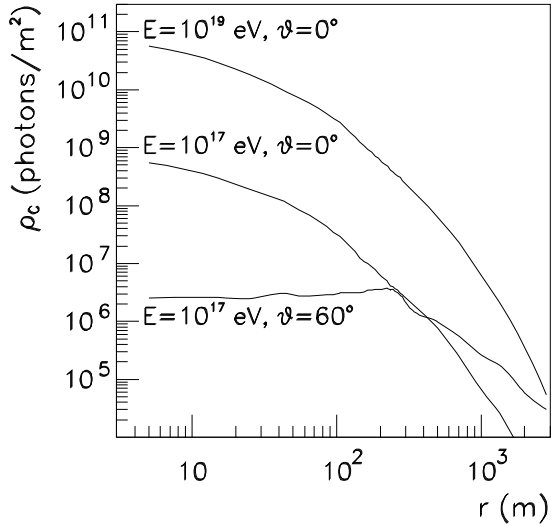}
}
\vspace{-0.8cm}
\caption{Cherenkov flux $\rho_C\ (\mbox{photons}/m^2)$ versus
 distance $r$ from the shower core for different CR energies and zenith angles.}
\label{ldf}
\end{figure}

\indent The flux of Extremely High Energy Cosmic Rays (EHE CRs) is
 very low: $\phi (E>10^{19}\ eV)\simeq 0.5\ km^{-2}yr^{-1}sr^{-1}$,
 and few hundreds of events have been recorded with energies above $10^{19}\
 eV$ \cite{rcdet}. Past and present experiments generally agree on the slope
 of the energy spectrum and on its absolute intensity below
 $E\simeq 4\cdot 10^{19}\ eV$; however, no
 firm conclusion can be drawn on the existence of events above
 the Greisen-Zatsepin-Kuz'min cut off in CR energy \cite{gzk}, on
 anisotropy of the arrival directions and correlation with
 cosmic point sources \cite{rcdir}, and on CR composition \cite{rccomp}. \\
\indent EHE cosmic rays are indirectly detected using several techniques:
 either through ground array experiments, which measure the lateral
 distribution of electrons and muons in the extensive air shower
 (EAS) using scintillation counters
 or water Cherenkov tanks; or through experiments sensitive to the UV
 photons emitted by nitrogen fluorescence generated in the atmosphere at the
 passage of the shower particles. \\
%
\indent Cherenkov light is also produced in the atmosphere by the electrically
 charged particles in the shower;
 this flux $\rho_C$ has a broad spatial distribution at sea level, with a
 lateral extent reaching several $km$ away from the EAS core for very
 energetic and inclined showers. Since $\rho_C$ roughly scales with
 CR energy, this light can be very intense: $\rho_C\simeq 10^{10}\
 \mbox{photons}/m^2$ for $10^{19}\ eV$ vertical CRs near the shower core
 \cite{kieda}, as shown in Fig.\ref{ldf}. \\
  The Cherenkov photons reach
 sea level as a plane wave, in a front with a typical duration of tens of
 $ns$ near the shower axis, and few $\mu s$ at several $km$ from the core.
 The spectral distribution of the Cherenkov light
 at sea level ranges between $300\ nm$ and $1500\ nm$. \\
\indent The properties of the Cherenkov pulse suggest the possibility to
 observe in a different way the highest energy CRs by means of low-cost
 semiconductor
 photodetectors, including solar panels \cite{kieda2}. Solar panels are
 composed of electrically connected solar cells, i.e. n-p junctions with
 a large surface area. Solar cells have a high quantum efficiency (QE),
 with a broad maximum between $600\ nm$ and $1000\ nm$, well matching the
 Cherenkov spectral distribution at sea
 level ($\langle QE\rangle _{\check{C}}\simeq 0.55\div 0.60$). The aim of
 this work is to
 experimentally investigate the viability of EHE cosmic rays observation
 using solar panels as air Cherenkov detectors. 
\section{EXPERIMENTAL SETUP}
\begin{figure}[tb]
\centerline{
\input{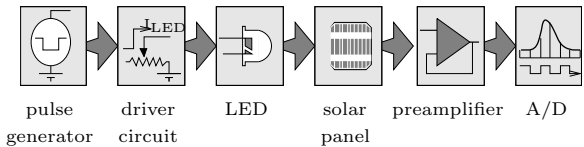}
}
\vspace{-0.8cm}
\caption{Block diagram of the experimental setup used
 to evaluate the response of solar cells/panels to light pulses.}
\label{setup}
\end{figure}

\begin{figure}[tb]
\centerline{
\epsfig{file=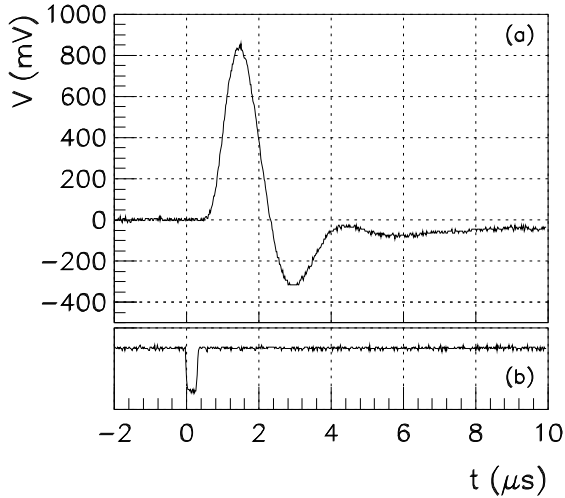}
}
\vspace{-0.8cm}
\caption{(a) Typical solar cell response with an ORTEC 142B preamplifier to
 a $\Delta t=300\ ns$ light flash; (b) square pulse driving the LED
 emission.}
\label{pulse}
\end{figure}

One of the main concerns is the evaluation of the response of single solar
 cells and
 panels to faint light pulses. Fig.\ref{setup} shows a schematic diagram
 of the experimental setup used for this purpose. \\
\indent In order to simulate the Cherenkov pulses, we used LEDs
 with fast response ($\simeq 15\ ns$) and good pulse-to-pulse reproducibility.
 Light flashes are produced through a driver circuit which allows both
 pulse duration and pulse intensity control.
 Suitable optical pulses were obtained, with $10^7\div 10^{10}$
 photons emitted per pulse, spread over a
 time duration of $100\ ns<\Delta t<1\ \mu s$. \\
\indent The photoelectric transient pulse produced by the cell/panel is
 decoupled from the continuous component through a
 capacitor or a pulse transformer and fed into a preamplifier; the resulting
 signal is then digitized and recorded. \\
%
\indent Particular attention has been devoted to the evaluation of the
 luminous power emitted by the
 LEDs; for this purpose we used different methodologies,
 both direct power measurements through large
 area calibrated sensors and measurements based on integrating spheres
 which are independent from light beam geometry. 
%
 We have tested monocrystalline, polycrystalline and amorphous Silicon solar
 cells with different
 active areas (from $4\ cm^2$ to $100\ cm^2$), and their grouping
 in panels and rows of panels through series or parallel connections.
 Tests
 included both commercial products and custom designed detectors in order
 to improve their transient response. \\ 
%
\indent The choice of the preamplifier to be used with solar cells/panels
 is a difficult task, and reflects
 the unusual properties of these photodetectors; both charge and
 voltage preamplifiers were tested. Charge preamplifiers are
 usually preferred with semiconductor detectors for their
 more stable voltage-to-charge gain, insensitive to the properties
 of the detector. 
 However, they are typically designed for capacitive detectors
 in the $pF$ to $nF$ range, a condition which does not
 apply to solar cells/panels. The net effect is that charge preamplifiers
 behave in a ``non ideal'' way.
 The ORTEC 142B charge preamplifier has
 shown to be a good candidate; a typical solar cell pulse obtained
 with this amplifier is shown in Fig.\ref{pulse}.
\section{SIGNAL CONSIDERATIONS}
In general, the speed of response of a photodiode is limited by the
 combination of three factors: carrier diffusion to the
 junction; drift time through the depletion region; effect of the charge
 storage in the depletion region, which can be expressed through the
 junction capacitance $C_d$.
 In the case of a solar cell,
 the main limitation is given by $C_d$, which
 is very high due to
 the low p-substrate resistivity and the large surface area.
 The cell may then be modelled as a photocurrent source in parallel to
 the junction capacitance; a cell shunt resistance $R_d$ has also to
 be considered to take into account the high leakage currents
 \cite{szebhatta}. \\
\indent We have verified that solar cells/panels behave as capacitive
 detectors by means of the pulse shape analysis of the photoelectric voltage
 observed through different load resistances; pulse shapes
 typical of a resistive charging of a capacitance were obtained.
\subsection{Voltage-to-charge linearity and gain} 
\begin{figure}[tb]
\centerline{
\epsfig{file=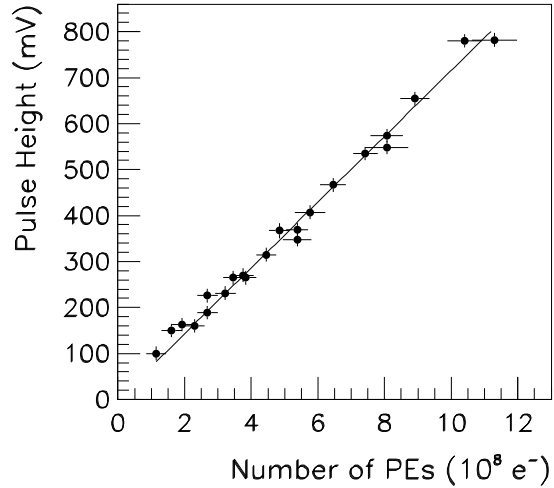}
}
\vspace{-0.8cm}
\caption{Pulse height versus number of photoelectrons for a $78\ cm^2$
 solar cell. The amplifier used is the ORTEC 142B.}
\label{1clin}
\end{figure}

\indent Fig.\ref{1clin}
 shows the linear relationship between the pulse
 height of the solar cell signal and the total photogenerated
 charge in the pulse,
 expressed as the number of photoelectrons (PEs). Signals
 in Fig.\ref{1clin} were obtained with different LEDs, varying
 intensities and pulse durations ($100\ ns<\Delta t<1\ \mu s$).
 From the slope
 of the linear fit we evaluated the voltage-to-charge
 gain. \\ 
\indent The measured voltage-to-charge ratio is dependent
 from $C_d$; moreover, it
 is lower by a factor of $10^2\div 10^3$ than what
 predicted by an ideal charge preamplifier,
 especially for the higher capacitance detectors tested
 (see Tab.\ref{tab:gainvscap}). \\
\begin{table}[tb]
\caption{Cell area, cell capacitance $C_d$ and
 measured gain for two cells of different areas and made from the same bulk
 material. The specified gain of the amplifier is $500\cdot 10^{-3}\ V/pC$.}
\label{tab:gainvscap}
\begin{tabular*}{7.5cm}{@{}rcc} \hline
Area $(cm^2)$ & $C_d\ (\mu F)$ & Gain $(10^{-3}\ V/pC)$ \\ \hline
 $9$  & $0.6\pm 0.1$   & $3.5\pm 0.3$ \\ 
 $96$ & $6.1\pm 1.4$   & $0.7\pm 0.2$ \\ \hline
\end{tabular*}
\end{table}
\indent Since the quantum efficiency is very similar
 for different cells, the detector capacitance is the main
 factor of merit for
 signal amplitude considerations, and must be reduced as much as
 possible. The use of higher resistivity substrates
 (even very large area PIN diodes) is a possible
 solution to further lower $C_d$, thereby
 increasing the preamp gain. 
\subsection{Solar cells grouping into modules}
\begin{figure}[tb]
\centerline{
\epsfig{file=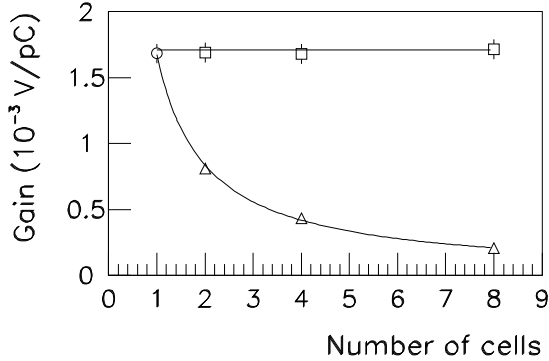}
}
\vspace{-0.8cm}
\caption{Voltage-to-charge gain for different electrical
 connections of identical cells. $\Box$: series connection;
 $\triangle$: parallel connection.}
\label{gainvscon}
\end{figure}

\indent An advantage of solar cells over other photodetectors
 is that they can easily be connected to give $1\ m^2$ or more
 active area, thus collecting more Cherenkov light. As shown in
 Fig.\ref{gainvscon}, we observed that
 a series connection of identical solar cells
 does not decrease the gain; then, increasing the number
 of series connected cells gives rise to a proportionally
 higher Cherenkov signal.
\subsection{Effect of background light}
\begin{figure}[tb]
\centerline{
\epsfig{file=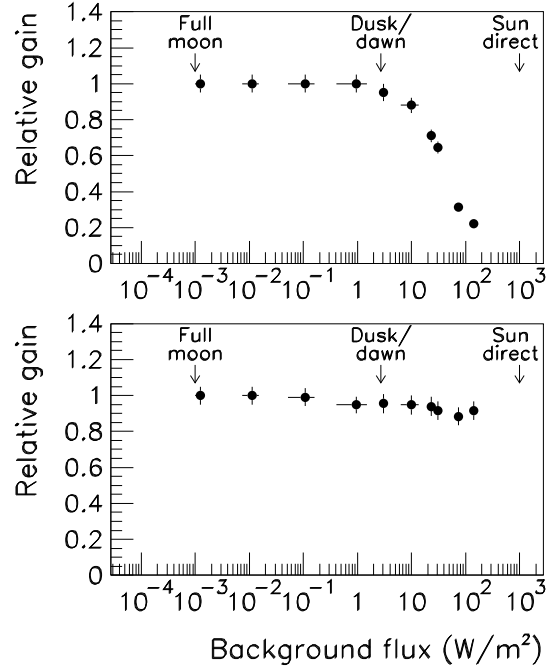}
}
\vspace{-0.8cm}
\caption{Pulse height degradation due to background light. Top: DC decoupling
 through capacitance; bottom: DC decoupling through low impedance pulse
 transformer.}
\label{bgr}
\end{figure}

\indent An important aspect concerning the duty cycle of solar panels
 as CR detectors is the gain degradation due to background light;
 in particular, the
 photovoltaic forward voltage due to ambient light increases the
 capacitance of the cells even further.
 Low impedance transformers allow to short out the DC photovoltaic
 voltage,
 and no significant signal degradation is observed even during daytime.
 This is not the case for solar panels working in
 open-circuit conditions, where the AC variation is obtained through a
 capacitive DC decoupler: gain degradation is observed even in dusk/dawn
 conditions (Fig.\ref{bgr}). However, in both cases the solar panel
 sensitivity may
 decrease during daytime due to additional shot-noise.
\section{DETECTOR NOISE AND SENSITIVITY}
\begin{figure}[bt]
\centerline{
\epsfig{file=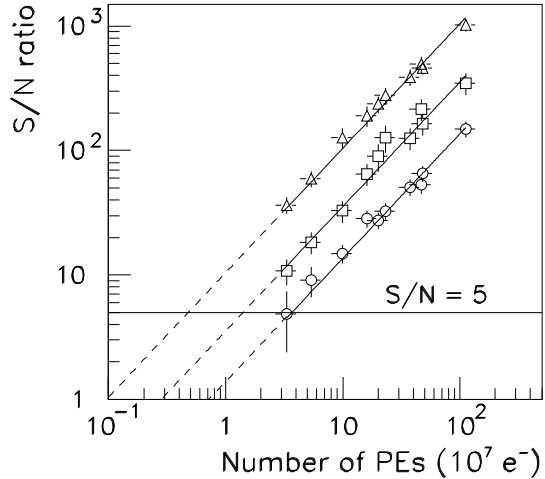}
}
\vspace{-0.8cm}
\caption{Signal-to-noise ratio versus the number of PEs for three
 different solar panels. $\bigcirc$: $36\cdot 96\ cm^2$ mono-Si cells;
 $\Box$: $32\cdot 32\ cm^2$ a-Si cells; $\triangle$: $36\cdot 9\ cm^2$
 mono-Si cells.}
\label{sn}
\end{figure}

%
For noise evaluation purposes, the solar cell equivalent circuit described
 above \cite{szebhatta} must be extended to include also series and
 parallel noise generators, which are dependent not only from the detector
 alone, but also from the preamplifier and shaper characteristics.
 In particular, the Equivalent Noise Charge (ENC) is lower for
 low junction capacitance, high shunt resistance cells, low series noise
 preamplifiers and long shaping times \cite{gatti}. Moreover, if the cell
 operates
 in the photovoltaic mode (no reverse bias applied) and under limited
 background light, the shot-noise contribution to ENC may be neglected.
 For our $10\times 10\ cm^2$ mono-Si cell and amplifier, we evaluated
 ENC$_{\mbox{rms}}\simeq 8\cdot 10^6\ e^-$, in good agreement with the
 observed value. \\
\indent Adding more series connected cells, RF pickup from the electrical
 connections in the panel becomes progressively the main source
 of noise. The fact that RF noise increases with detector area
 was clearly verified from observations performed on the $30\ m^2$ rows of
 an experimental photovoltaic plant operated by ENEA at Manfredonia (Italy).
 As single panels are used, proper RF shielding is obtained using Faraday
 cages; in this way, we do not observe any significant variation in ENC in
 increasing the number of series connected cells. \\
 \indent Fig.\ref{sn} shows the
 signal-to-noise ratio in dark conditions plotted versus the number of PEs,
 for three different panels. These are a mono-Si $0.35\ m^2$ module
 (Italsolar 36 MS-CE), an a-Si $0.10\ m^2$ module (Solarex SA-5), and a
 special mono-Si $0.032\ m^2$ module manufactured from EUROSOLARE to test
 smaller area cells.
Considering a trigger threshold of $5\sigma$, the PE sensitivity per
 unit area is $\simeq 10^8\ e^-/m^2$, similar for the three panels and
 comparable to the charge per unit area produced in the panels by the
 Cherenkov flux of $10^{17}\ eV$ vertical CRs near the shower core
 (see Fig.\ref{ldf}).   
\section{SIMULATION OF CR DETECTION CAPABILITIES}
To give a quantitative
 idea of CR detection capabilities of solar panel detectors, we have
 considered both a small array for $\sim 10^{17}\div 10^{18}\ eV$ CRs,
 and a larger one for the
 highest energy CRs. Their characteristics are given in Tab.\ref{tab:arrays}.
 \\
\begin{table}[tb]
\setlength{\tabcolsep}{1.5pc}
\newlength{\digitwidth} \settowidth{\digitwidth}{\rm 0}
\catcode`?=\active \def?{\kern\digitwidth}
\caption{Two possible arrays for CR detection.}
\label{tab:arrays}
\begin{tabular*}{7.5cm}{@{}c@{\extracolsep{10pt}}c@{}c} \hline
                   & A &  B  \\ \hline
Number of det. & $16$ & $256$ \\ 
Det. spacing & $50\ m$ & $500\ m$ \\ 
Array config. & \multicolumn{2}{@{}c}{rectangular grid} \\ 
Det. type      & \multicolumn{2}{@{}c}{mono-Si panel
 (36 series cells)} \\ 
Det. area      & \multicolumn{2}{@{}c}{$0.35\ m^2$}  \\ 
Det. orientation & \multicolumn{2}{@{}c}{flat, horizontal} \\ 
Det. sensitivity & \multicolumn{2}{@{}c}{$3.6\cdot 10^7\ e^-$}  \\ \hline
\end{tabular*}
\end{table}  
\indent The predictions are based on Cherenkov lateral and
 spectral distributions from a modified MOCCA code
 including Cherenkov light production and attenuation in atmosphere
 \cite{kieda}.
 CRs are generated as primary protons with isotropic arrival directions
 and with energies according to the observed CR flux
 \cite{watson}. Knowing the
 Cherenkov flux reaching the detectors, the panels quantum efficiency
 weighted on the Cherenkov spectrum and the
 effective area of the
 detector seen by the Cherenkov wave plane, one can deduce the photogenerated
 charge in the detectors.
\subsection{Event rate and energy spectrum}
\begin{figure}[tb]
\centerline{
\epsfig{file=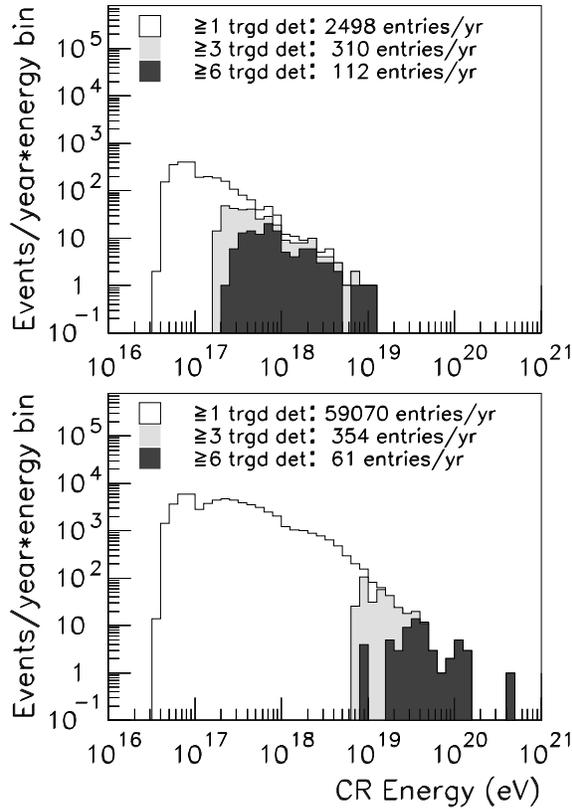}
}
\vspace{-0.8cm}
\caption{Energy spectrum of detectable CRs with the
 array A (top) and the array B (bottom). The statistics refers
 to one year of simulated data-taking. The total number of events for
 different trigger conditions is also given.}
\label{enspectr}
\end{figure}

 Fig.\ref{enspectr} shows the event rate and the
 event energy distribution for the two arrays of Tab.\ref{tab:arrays},
 for different thresholds on the number of triggered detectors.
 Both arrays give $\simeq 1$ event/day when three triggered detectors
 are required. This rate may decrease by a factor of $2\div 3$ when
 meteorological conditions and background light are taken into account.
\subsection{Sky coverage}
\indent We also evaluated the solid angle of acceptance for the
 array B of Tab.\ref{tab:arrays}.
 This array should give almost uniform sky coverage up to $\simeq 60^o$.
 For greater zenith angles very few CRs are observable with horizontal
 detectors, because of the lower Cherenkov flux, the smaller detector
 effective area and the higher reflectivity of the panels coating.
 It has been proposed to use several panels oriented towards
 different regions of the sky for each detector, in order to
 reconstruct the CR arrival direction even with no timing information
 \cite{suson}.
 In this way a complete sky coverage could also be obtained.
\section{CONCLUSIONS}
The response of solar panels to light pulses was investigated
 both theoretically and experimentally.
 These devices show excellent quantum efficiency and linearity;
 satisfactory sensitivities have been reached ($\simeq 10^8\ e^-/m^2$).
 Optimization of solar panels for Cherenkov light detection is
 possible, but even commercial modules seem to be adequate. \\
\indent A Monte-Carlo study was performed in order to predict CR
 detection capabilities of possible arrays. The results indicate that
 the technique may be tested at $\sim 10^{17}\
 eV$ energies, allowing an evaluation of
 the accuracy in the reconstruction of shower parameters.
 In conclusion, cost-effective
 solar panels could be strong candidates for the detection of the EHE CRs.
\section*{ACKNOWLEDGEMENTS}
The authors would like to express their gratitude to D.~B.~Kieda
 for the original idea on this detection technique and for several data
 used in the Monte-Carlo calculations. We acknowledge M.~Bruno, G.~Martinelli
 and G.~Tomassetti for their suggestions and their kind supply of electronics
 instrumentation.
 We are also indebted to R.~Peruzzi from EUROSOLARE (Nettuno, Italy) for
 providing us with solar panels specifically manufactured for this work. 
\end{document}